\documentclass[times,10pt]{article}
\usepackage{amssymb,amsmath}
\usepackage{graphicx}
\usepackage{epstopdf}

\begin{document}
\centerline{\bf Exact Statistical Thermodynamics of  the Pseudospin-1 System on the Diced Lattice}\vskip .2in

%\centerline{September 4, 2020}\vskip .2in  

\centerline{M.L. Glasser}
\centerline{Department of Physics}
\centerline{Clarkson University}
\centerline{Potsdam, NY 13699, USA}\vskip .1in
\vskip .3in\centerline{Norman J. M. Horing}
\centerline{Department of Physics}
\centerline{Stevens Institute of Technology}
\centerline{Hoboken, NJ 07030, USA}\vskip .2in
\vskip .1in

\centerline{\bf ABSTRACT}
\vskip .1in

\begin{quote}
In this work we analyze the thermodynamic properties of the pseudospin-1 Hamiltonian on the two-dimensional {\cal{T}}
-3 or Diced Lattice. Starting from the Partition function, we obtain the Grand ensemble thermodynamic potential, entropy and specific heat exactly and in the degenerate and non-degenerate regimes.\end{quote}
\vskip .4in\noindent
PAC:73.20-r, 73.43-r


\newpage

\section{ Introduction}
\vskip .1in

This work addresses the thermodynamic properties of the Fermion pseudospin-1 system whose Hamiltonian  is
$${\cal{H}}=(\hbar v/\sqrt{2})\left[\begin{array}{ccc}
0&K_-&0\\
K_+&0&K_-\\
0&K_+&0
\end{array}\right],\eqno(1.1)$$ which was  introduced by D. Bercioux, et al.[1] in 2009, where $v$ is the effective Fermi velocity, $k_{\pm}=k_x\pm i k_y$ and $\vec{K}$ is the 2D- crystal momentum.This system has attracted attention recently, particularly by Malcom and Nicol [2] who studied its electronic polarizability and related properties. This system falls into the class of Dirac materials, which includes group IV Dichalcogenides[3], Topological Insulators[4], Silicene[5] and, most notably, Graphene[6-11].

In Section 2,  we derive the exact electronic Thermodynamic Potential, Entropy, Chemical Potential and Specific Heat, which can all be expressed in closed form, and in Section 3 we discuss the behavior of these quantities  in the degenerate and non-degenerate regimes which may be relevant to possible experimental conditions. The results are summarized and discussed in Section 5.

\vskip .2in
\section{ Calculations}\vskip .1in

In the body of this work we set $\hbar=1$ and introduce the  dimensionless momentum $\vec{k}=\vec{K}/K_m$, where $K_m$ is a cut-off introduced to restrict the band structure to the relativistic region where  the dispersion is linear. We also introduce the characteristic energy  $a=\hbar vK_m$ corresponding to the highest energy in the linear portion of the spectrum. The partition function, worked out in the Appendix, is
$$Z(s)=\frac{K_m^2}{\pi}[\frac{1}{4}+\frac{1}{a^2s^2}-\frac{\cosh(as)}{a^2s^2}+\frac{\sinh(as)}{as}].\eqno(2.1)$$

To proceed, we note the Wilson-Sondheimer formula [13,14] giving the Grand Thermodynamic Potential 
$$\Omega=-k_BT\sum_{\{\alpha\}}\ln[1+e^{-\beta(E_{\alpha}-\mu}]=-\frac{\beta}{4}\int_0^{\infty}dt\frac{z(t)}{\cosh^2[\frac{\beta}{2}(t-\mu)]},\eqno(2.2)$$
( $k_B$ being Boltzmann's constant)
where  the sum is over all the energy levels, $T$ is absolute temperature, $\beta=1/k_BT$, $\mu$ is the Chemical Potential and 

$$z(t)=\int_{c-i\infty}^{c+i\infty}\frac{ds}{2\pi is^2}e^{st} Z(-s)=$$

$$z(t)=\frac{K_m^2}{12\pi a^2}\left[t^3+6a^2t+2a^3-\left(t^3-3a^2t+2a^3\right)\theta(t-a)\right] ,\eqno(2.3)$$
Here $\theta(z)$ denotes the Heaviside step function and, due to the prefactor $K_m^2$, $\Omega$ and subsequent quantities refer to unit area. We point out that, apart from the system parameters $K_m$ and $a$, $\Omega$ is a function of the state variables $T$ (or $\beta$) and $\mu$ (or $N$) since the volume (i.e. area) is fixed. Once the density is specified, as in dealing with a specific sample, $\mu$ is itself a function of $T$.
Since (2.3) may be unfamiliar, some details are provided in the appendix.
The function $z(t)$, is basically a cubic polynomial in $t$ and  behaves nearly linearly over its range.  Not surprisingly it is found that the integral resulting from inserting it into (2.3) can be evaluated exactly to give
 $$\Omega=-\frac{K_m^2}{12\pi\beta}\{\beta(3\mu-a)+a\beta\tanh(\beta\mu/2)+9\ln(1+e^{\beta\mu})$$
 $$-\frac{6}{(a\beta)^2}[a\beta Li_2(-e^{\beta(a-\mu)})-Li_3(-e^{\beta(a-\mu)})+Li_3(-e^{-\beta\mu})]\}\eqno(2.4)$$
where
$$Li_n(z)=\sum_{k=1}^{\infty}\frac{z^k}{k^n}\eqno(2.5)$$
is the polylogarithm[15] (See Appendix E)..
\vskip .1in

The chemical potential is  the Lagrange coefficient associated with the constraint that the system contain $N$ particles and is closely related to the energy necessary to add or remove one particle . Since in most situations the areal density  $n$ is fixed, it is useful to know how $\mu$ is related to $n=-\partial\Omega/\partial\mu$. From (2.4) one finds

\newpage
$$n=\frac{K_m^2}{12\pi}\left\{12-\frac{9}{1+e^{\beta\mu}}+\frac{a\beta}{\cosh(\beta\mu)+1}-\frac{6}{a\beta}\ln[1+e^{\beta(a-\mu)}]-\frac{6}{a^2\beta^2}[Li_2\left(-e^{\beta(a-\mu)}\right)-Li_2\left(-e^{-\beta\mu}\right)]\right\}
\eqno(2.6)$$
\vskip .1in
Next, the entropy $S=k_b\beta^2\partial\Omega/\partial \beta$, is given by
$$\frac{24\pi a^2}{K_b K_m^2}S=$$
$$ 6a\{2(\mu-a)\ln[1+e^{\beta(a-\mu)}]+3a\ln[1+e^{\beta\mu}]+3a\ln[1+e^{-\beta\mu}]\}-\frac{3\beta\mu}{1+e^{-\beta\mu}}-a^3\beta^2\mu{\rm sech}^29\beta\mu/2)\eqno(2.7)$$
$$+12\beta^{-1}(\mu-3a)Li_2(-e^{\beta(a-\mu)})+\frac{12}{\beta^2}\left[2Li_3(-e^{\beta(a-\mu)})-\beta\mu Li_2(-e^{-\beta\mu})-2Li_3(-e^{-\beta\mu})\right]$$

   \section{ Degenerate Limit }\vskip .1in
   
   As $T\rightarrow0$ the system Fermions condense into the lowest states  up to the Fermi energy, which in the limit coincides with $\mu$. In this case it is convenient to write (2.5) 

 $$\Omega=\int_0^{\infty} dt f_0'(t) z(t)=-\frac{z(0)}{1+e^{\beta\mu}}-\int_0^{\infty}dt\;  f_0(t) z'(t)\eqno(3.1)$$
 where
 $$f_0(t)=\frac{1}{1+e^{\beta(t-\mu)}}.\eqno(3.2)$$
 Now, the last integral is, after the change of variable $\beta(t-\mu)\rightarrow t$ 
 $$\frac{1}{\beta}\int_{-\beta\mu}^{\infty}dt\; \frac{z'(\mu+t/\beta)}{e^t+1}=\frac{1}{\beta}\left[\int_0^{\beta\mu}z'(\mu-t/\beta)\left(1-\frac{1}{e^t+1}\right)+\int_0^{\infty}\frac{z'(\mu+t/\beta)}{e^t+1}dt\right]\eqno(3.3)$$
 Since, in the degenerate limit, $\beta\mu\rightarrow\infty$,
 $$\Omega=-\frac{z(0)}{1+e^{\beta\mu}}-z(0)+z(\mu)-\frac{1}{\beta}\int_0^{\infty}\frac{z'(\mu-t/\beta)-z'(\mu+t/\beta)}{e^t+1}dt.\eqno(3.4)$$
 Now,
 $$z'(\mu\pm t/\beta)=z'(\mu)\pm \frac{t}{\beta} z''(\mu)+\frac{t^2}{2\beta^2}z'''(\mu)+\cdots\eqno(3.5)$$
 so
$$\Omega_{deg}=\left(\frac{2+e^{\beta\mu}}{1-e^{\beta\mu}}\right)z(0)-z(\mu)+\frac{z''(\mu)}{2\beta^2}\int_0^{\infty}\frac{t}{1+e^t}dt+\cdots.\eqno(3.6)$$
 Therefore, since we can ignore $e^{-\beta\mu}$, for any Fermi system
$$\Omega_{deg}=Const-z(\mu)-\frac{\pi^2}{6}z''(\mu)(k_bT)^2+O((k_b T)^4).\eqno(3.7)$$
Since the Fermi temperature is on the order of kiloKelvins, it is this regime which applies to most experimental situations.. Also, since $a$ is roughly the maximum energy level, we can assume $\mu<a$,in which case one has
from (2.3) and (3.7)
$$\Omega_{deg}=-\frac{K_m^2}{12\pi a^2}\left[\mu^3+6a^2\mu+2a^3\right] .\eqno(3.8)$$
 More details are provided in[16].
\vskip .2in

\section{Non-degenerate limit}\vskip .1in

To investigate the high temperature, low density behavior, we rewrite (3.1) in the form
$$\Omega=\int_{c-i\infty}^{c+i\infty}\frac{ds}{2\pi i}\frac{Z(-s)}{s^2}\int_0^{\infty}dt \; e^{st} f_0'(t).\eqno(4.1)$$
In this case. it is appropriate to write
$$f_0'(t)\approx -\beta e^{\beta\mu} e^{-\beta t}.\eqno(4.2)$$
and, therefore, the right hand side of (4.1) is simply the inverse Laplace transform of a Laplace transform yielding
$$\Omega_{nd}=-\frac{1}{\beta}e^{\beta\mu}Z(\beta).\eqno(4.3)$$ 
where $Z(\beta)$ is given in (2.4).
Consequently, the density is
$$n=-\frac{\partial\Omega_{nd}}{\partial \mu}=e^{\beta\mu}Z(\beta)=-\beta\Omega_{ng}\eqno(4.4)$$
Note that this relationship is valid for any Fermi system.\vskip .2in

For further examination of this region, see the previous paper in this series[16].

\section{Discussion}\vskip .1in

The thermodynamic quantities obtained so far are all expressed in teems of the chemical potential, $\mu$, and are of little use without knowing how $\mu$ depends on the electron density $n$. Except in the degenerate regime, where this has been carried out in [16], one must solve (2.6) for $\mu$, a formidable task. Therefore, we shall attempt to proceed numerically by specifying $n$, $a$ and looking at null contours of the $n$-$\beta$-$\mu$ plots. In addition, in order to avoid the complication of an additional parameter, we introduce the scaled density $\nu=12\pi n/K_m^2$  and solve 
$$\nu
-\{12-\frac{9}{1+e^{\beta\mu}}+\frac{a\beta}{\cosh(\beta\mu)+1}$$
$$-\frac{6}{a\beta}\ln[1+e^{\beta(a-\mu)}]-\frac{6}{a^2\beta^2}[Li_2\left(-e^{\beta(a-\mu)}\right)-Li_2\left(-e^{-\beta\mu}\right)\}=0\eqno(5.1)$$
for $\mu$ vs $\nu$. Because the behavior is not very sensitive to $a$, we consider only the two values $a=10$ and $a=100$ atomic units ($\hbar=2m=1$, $e^2=2$). A reasonable temperature range is $4K<T<400K$ corresponding roughly to $3\cdot10^3<\beta<4\cdot10^5$. We arbitrarily assume $0<\nu<1000$. From the results shown in Figs 1,2 it appears that the chemical potential is virtually independent of temperature, rises steeply at low values of $\nu$, but, indicating density-induced  phase transition near $\nu=10$, remains nearly constant at $\mu=a$ at all higher densities

\begin{figure}[htbp]
    \centering
    \includegraphics[width=5cm]{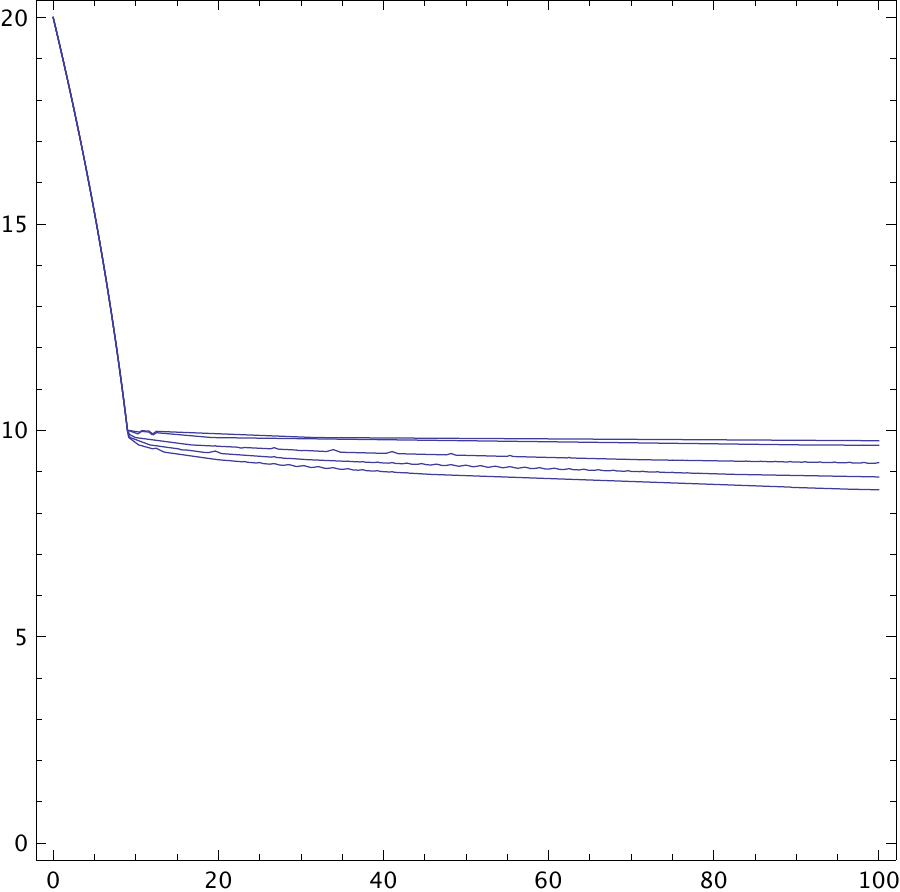}
    \caption{$\mu$ vs $\nu$ for $a=10$ at $T=4,10,100$ and $300K$]}
   \label{fig1}
\end{figure}

\begin{figure}[htbp]
    \centering
    \includegraphics[width=5cm]{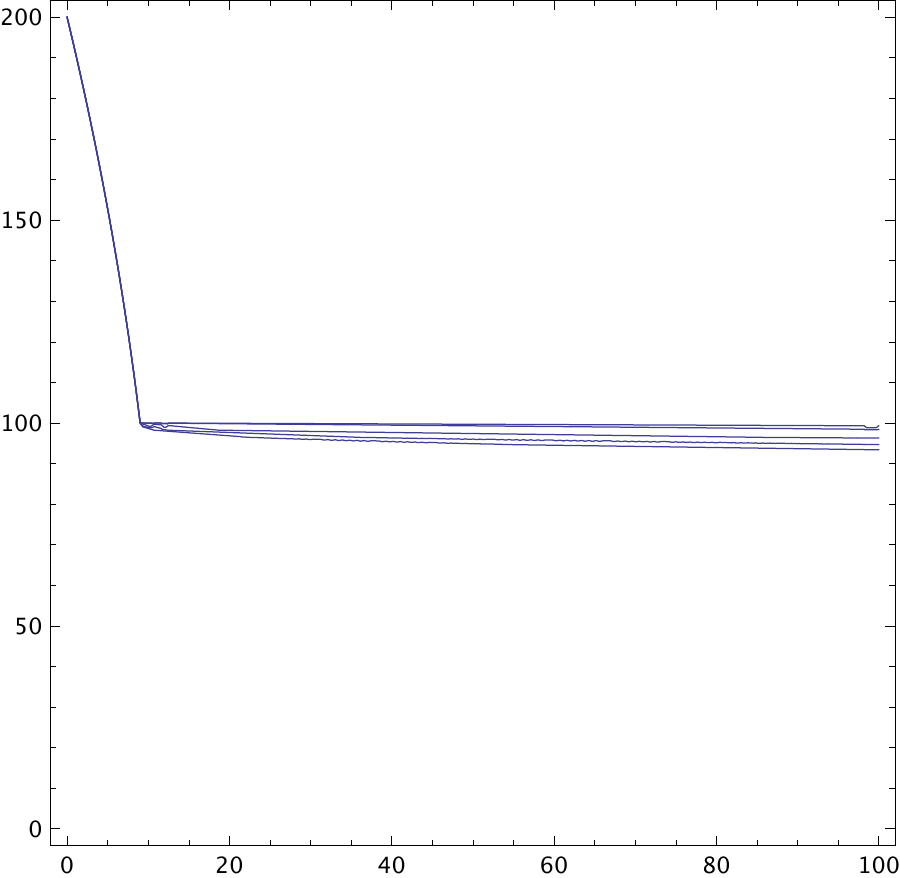}
    \caption{$\mu$ vs $\nu$ for $ a=100$ at $T=4,10,100$ and $400K$}
   \label{fig2}
\end{figure}

Next, we examine the thermodynamic potential (2.4). Introducing the scaled version $\bar{\Omega}=-12\pi\Omega/K_m^2$, one has
$${\bar \Omega}=A(a,\beta,\mu)+B(a,\beta,\mu)$$
$$A(a,\beta,\mu)=3\mu-a(1-\tanh(\beta\mu/2))+\frac{9}{\beta}\ln(1+e^{\beta\mu})\eqno(5.2)$$
$$B(a,\beta,\mu)=-\frac{6}{(a\beta)^2}[a Li_2(-e^{\beta(a-\mu)})-\frac{1}{\beta}\{Li_3(-e^{\beta(a-\mu)})-Li_3(-e^{\beta\mu})\}].$$
Numerically, we find that $A$ and $B$, and so ${\bar \Omega}$ are nearly temperature independent over the range $4K<T<400K$ and that $A$ is nearly independent of $a$ over tis range. Consequently, quantities related to temperature derivatives of the thermodynamic potential, such as the entropy and specific heat, will be small, as seen for the dimensionless entropy if Fig.3.. Of course, at higher temperatures where states no longer described by (1.1) are occupied, the situation will be different.\vskip .1in

The scaled entropy 
$${\bar \Sigma}=\frac{12\pi S}{k_B K_m^2}$$
is sketched in Fig.3

\begin{figure}[htbp]
    \centering
    \includegraphics[width=5cm]{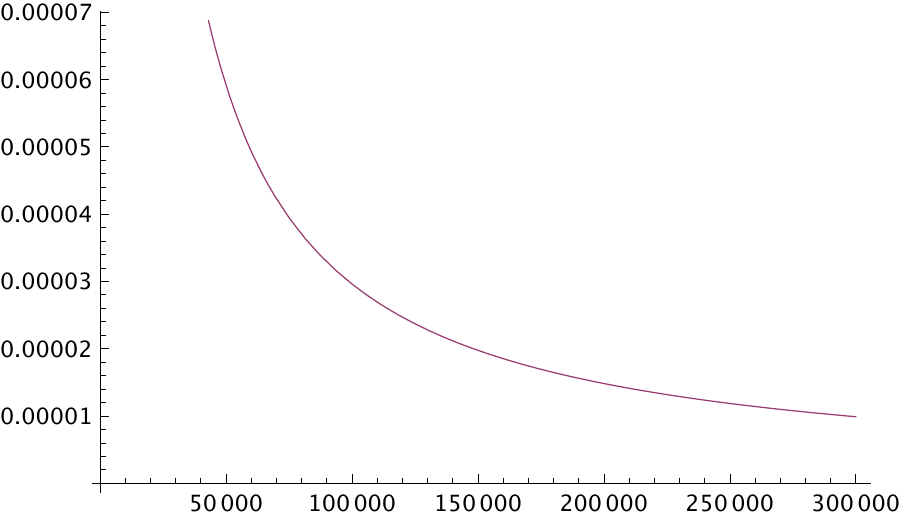}
    \caption{${\bar \Sigma}$ vs inverse temperature $\beta$ for $400K>T>4K$. Here $a=10$, $\nu=20$]}
   \label{fig3}
\end{figure}

\vskip .2in
The Specific Heat is
$$C_V=-T\frac{\partial S}{\partial T}=$$
$$\frac{K_BK_m^2}{12\pi a^2} (\frac{4 a^3 b^3 m^2}{(e^{b m}+1)^3}+\frac{a^2 b^2 m (a (4-6 b m)+9
   m)}{(e^{b m}+1)^2}+\frac{a^2 b^2 m (2 a (b m-2)-9 m)}{e^{b m}+1}$$
   $$+6
   (3 a^2-4 a m+m^2) \log (e^{b (a-m)}+1)  \frac{6 a b e^{a b} (a-m)^2}{e^{a b}+e^{b m}}-6 m^2 \log
   (e^{-b m}+1))$$
   $$+\frac{12 (b (3 a-2
   m) {Li}_2(-e^{b (a-m)})-3 {Li}_3(-e^{b (a-m)})+2 b m
   {Li}_2(-e^{-b m})+3 {Li}_3(-e^{-b
   m}))}{b^2}$$

   and  is shown in Fig.4 for $a=10$, $100K> T >4K$

   \begin{figure}[htbp]
    \centering
    \includegraphics[width=5cm]{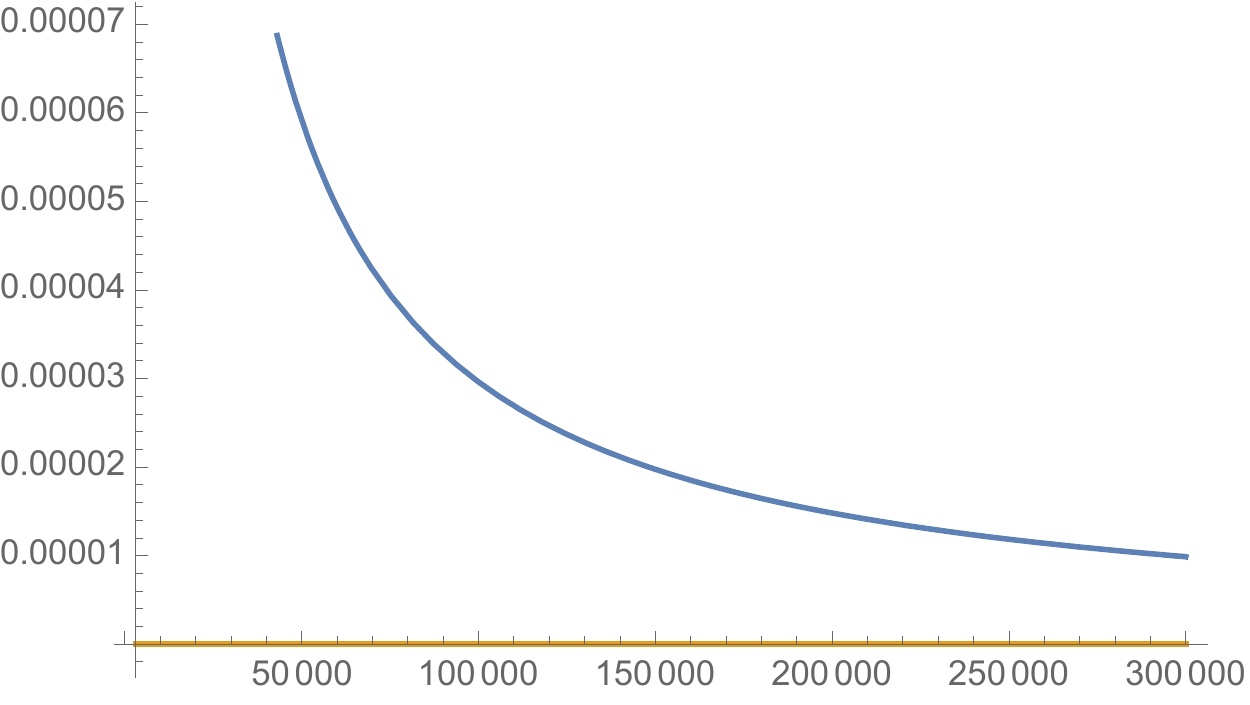}
    \caption{$C_V$ vs inverse temperature $\beta$ for $400K>T>4K$. Here $a=10$, $\nu=10$]}
   \label{fig4}
\end{figure}

\newpage

\centerline{\bf References}\vskip .1in

\noindent
[1] D. Bercioux,D.F. Urban, H. Grabert and W. Ha\"usler, Phys. Rev{{\bf A80},063683 (2009)

\noindent
[2]  J.D. Malcolm and E.J.Nicol, Phys. Rev.{\bf B93},165433 (2016)

\noindent
[3]  {\it Applications of Dichalcogenides S, Se, Te},[Ed.  G.K. Ahluwalia,, Springer (2017)]

\noindent
[4]  S.Q. Shen,{\it Topological Insulators},[ Springer (2012)].

\noindent
[5] M.J.S. Spencer,{\it Silicene},[Springer (2016)]

\noindent
[6]  M. Katsnelson,{\it Graphene, Carbon in Two-Dimensions},[ Cambridge UP (2012)]

\noindent

[7] H. Aoki and M.S. Dresselhaus,{\it Physics of Graphene},[Springer (2013)]

\noindent
[8] E.L. Wolf,{\it Grahene, A New Paradigm in Condensed Matter and Device Physics },[Oxford UP (2013)]

\noindent
[9] Y.O. Wehling, A.M. Black-Shaffer and A.V. Balatsky, ArXiv 1405.5774v1 (2014).

\noindent
[10]  J. Weng, S. Deng, Zhongfan Liu and Zhirong Liu, {\it The Rare 2D Materials with Dirac Cones},[ National Science Review 2, 22-39 (2015)]

\noindent
[11] N.J.M. Horing, Trans. Roy. Soc {\bf A368},5525 (2010).

\noindent
[12] N.J.M. Horing,{\it Quantum Statistical Field Theory},[ Oxford UP (2017)]

\noindent
[13] E.H. Sondheimer and A.H. Wilson, Proc. Roy. Soc.LOndon {\bf 210}, 273 (1951).

\noindent
[14] A.H. Wilson {\it Theory of Metals},[Cambridge UP , Sec 6.6 (1965)]

\noindent
[15] NIST on-line Handbook of Mathematical Functions {\it https://dlmf.nist.gov/25.12\#ii}.

\noindent
[16] N.J.M. Horing, M.L. Glasser and J. D. Mancini, {\it Statistical Thermodynamics of te Diced Lattice} (unpublished).

\newpage
\centerline{\bf Appendix}

-
}\centerline{\bf A: Partition Function}\vskip .2in

  Let $\vec{K}=K_m\vec{k}$ and $a=\hbar v K_m$
 $$H=\frac{a}{\sqrt{2}}  \left[\begin{array}{ccc}
  0&k_-&0\\
  k_+&0&k_-\\
  0&k_+&0\end{array}\right] ={\cal{U}}^{-1}\left(\begin{array}{ccc}
  0&0&0\\
  0&ak&0\\
  0&0&-ak
  \end{array}\right){\cal{U}}\eqno(A-1),$$
  where the unitary matrix $\cal{U}$ consists of the normalized eigenvectors of  $H$, has eigenvalues $\omega=0, \; \pm ak.$ Then
   the partition function is
 $$Z(s)=\sum_{\{\omega\}} e^{-s\omega } =\frac{K_m^2}{2\pi}  \int_0^1 dk\; k[1+2\cos(ask)]\eqno(A-2)$$
   $$=\frac{K_m^2}{\pi}\left[\frac{1}{4}+\frac{1}{a^2s^2}(1+as\sinh as-\cosh as)\right],$$
   since
   $\int dx \; x\cosh(ax)=[(ax)\sinh(ax)-\cosh(ax)]/a^2$.

\vskip .2in
\centerline{\bf B: Derivation of (2.2)}\vskip .1in

Since any function is the inverse Laplace transform of its Laplace transform,
$$ f(E)=k_BT\ln[1+e^{\beta(E-\mu)}]=\int_{c-i\infty}^{c+i\infty}\frac{ds}{2\pi i}e^{Es}\int_0^{\infty}dt e^{-st}f(t).\eqno(B-1)$$
From (2.5),
$$\Omega=\int_{c-i\infty}^{c+i\infty}\frac{ds}{2\pi i} Z(-s)\varphi(s)\eqno(B-2)$$
where $\varphi(s)$ is the second integral in (B-1). By multiplying and dividing the integrand of (B-2) by $s^2$ and noting that  $s^2\varphi(s)$ is the Laplace transform of $ f''(t)=-(\beta/4){\rm sech}^2(\left[\frac{\beta}{2}(t-\mu)\right]$.
With $z(t)$ defined in (2.6)  one has the Sondheimer-Wilson formula. \vskip .2in

\centerline{\bf C: Derivation of (2.3)}\vskip .1in

$$ Z(-s)=Z(s)=\frac{K_m^2}{\pi}\left[\frac{1}{4}+\frac{1}{p^2}+\frac{\sinh(p)}{p}-\frac{\cosh(p)}{p^2}\right]\eqno(C-1)$$
where $p=as$. Hence
$$\frac{Z(s)}{s^2}=\frac{K_m^2a^2}{\pi}\left[\frac{1}{4p^2}+\frac{1}{p^4}+\frac{e^p}{2}\left(\frac{1}{p^3}-\frac{1}{p^4}\right)-\frac{e^{-p}}{2}\left(\frac{1}{p^2}+\frac{1}{p^4}\right)\right].\eqno(C-2)$$
Thus,
$$z(t)=\frac{K_m^2a^2}{\pi}\left[L^{-1}_{t/a}[\frac{1}{4p^2}+\frac{1}{p^4}]+\frac{1}{2}L^{-1}_{(t+a)/a}[\frac{1}{p^3}-\frac{1}{p^4}]-\frac{1}{2}L^{-1}_{(t-a)/a}[\frac{1}{p^3}+\frac{1}{p^4}]\right]\eqno(C-3)$$
since an exponential factor just shifts the argument of the inverse Laplace transform, which is $t/a$ after changing the variable of integration from  s to p. Finally, using
 $$L^{-1}_x[p^{-n}]=\frac{1}{(n-1)!} x^{n-1}\theta(x)\eqno(C-4)$$
one gets 

$$z(t)=\frac{K_m^2}{12\pi a^2}\left[t^3+6a^2t+2a^3-\left(t^3-3a^2t+2a^3\right)\theta(t-a)\right]\eqno(C-5)$$

\centerline{\bf D: Calculation of $\Omega$}\vskip .1in

By inserting (2.3) into (2.4) one has a linear combination of the four elementary integrals

$$\int_0^{\infty }  \rm{sech}^2(\frac{1}{2} \beta (t-\mu)) \, dt=\frac{4 e^{\beta\mu}}{\beta (e^{\beta \mu}+1)}$$
$$\int_0^{\infty } t \,  \rm{sech}^2(\frac{1}{2} \beta (t-\mu)) \, dt=\frac{4 \ln (e^{\beta \mu}+1)}{\beta^2}$$

$$\int_0^{\infty } t^2\,  \rm{sech}^2(\frac{1}{2} \beta (t-\mu)) \, dt=-\frac{8 \rm{Li}_2(-e^{\beta \mu})}{\beta^3}\eqno(D-1)$$

$$\int_0^{\infty } t^3\, \rm{sech}^2(\frac{1}{2} \beta (t-\mu)) \, dt=\frac{4 (\beta^3 \mu^3-6 \rm{Li}_3(-e^{-\beta \mu})+\pi ^2 \beta \mu)}{\beta^4}$$

wihich follow from the elementary integrals

$$
\begin{array}{cc}
n&\int \frac{t^n}{\cosh^2(a x)}dx\\
&\\
 0 & \frac{\tanh (a x)}{a} \\
 1 & \frac{a x \tanh (a x)-\log (\cosh (a x))}{a^2} \\
 2 & \frac{{Li}_2\left(-e^{-2 a x}\right)+a x \left(-a x+a \tanh (a x) x-2 \log
   \left(1+e^{-2 a x}\right)\right)}{a^3} \\
 3 & \frac{2 a^2 \left(-a x+a \tanh (a x) x-3 \log \left(1+e^{-2 a x}\right)\right)
   x^2+6 a {Li}_2\left(-e^{-2 a x}\right) x+3 {Li}_3\left(-e^{-2 a
   x}\right)}{2 a^4}
\end{array}\eqno(D-2)
$$
\vskip .2in
\centerline{\bf The Polylogaritm}\vskip .1in

\noindent
Fpr $z$ small
$$Li_2[-e^{-z})\approx -\frac{\pi^2}{12}+\ln(2)z-\frac{z^2}{4}$$
$$Li_3(-e^{-z})\approx -\frac{3}{4}\zeta(3)+\frac{\pi^2}{12}z-\frac{1}{2}\ln(2)z^2$$

For large $z$
$$Li_2(-e^{-z})\approx -e^{-z}+\frac{1}{4}e^{-2z} $$
$$Li_3(-e^{-z})\approx -e^{-z}+\frac{1}{8}e^{-2z}.$$
\vskip .2in

\end{document}